\documentclass[aps,prb,preprint,groupedaddress,showpacs]{revtex4}

\usepackage{graphicx}
\usepackage{dcolumn}
\usepackage{bm}
\usepackage{subfigure}

\begin{document}


\title{Aspect-Ratio Scaling of Domain Wall Entropy for the 2D $\pm J$
Ising Spin Glass}


\author{Ronald Fisch}
\email[]{ron@princeton.edu}
\affiliation{382 Willowbrook Dr.\\
North Brunswick, NJ 08902}


\date{\today}

\begin{abstract}
The ground state entropy of the 2D Ising spin glass with +1 and -1
bonds is studied for $L \times M$ square lattices with $L \le M$ and
$p$ = 0.5, where $p$ is the fraction of negative bonds, using
periodic and/or antiperiodic boundary conditions.  From this we
obtain the domain wall entropy as a function of $L$ and $M$.  It is
found that for domain walls which run in the short, $L$ direction,
there are finite-size scaling functions which depend on the ratio $M
/ L^{d_S}$, where $d_S = 1.22 \pm 0.01$.  When $M$ is larger than
$L$, very different scaling forms are found for odd $L$ and even
$L$. For the zero-energy domain walls, which occur when $L$ is even,
the probability distribution of domain wall entropy becomes highly
singular, and apparently multifractal, as $M / L^{d_S}$ becomes
large.

\end{abstract}


\maketitle

\section{Introduction}

Recently, a series of unexpected results and
conjectures\cite{JLMM06,AHHM06,Fis06b,Fis07} has provided a new
perspective on the nature of the two-dimensional (2D) Ising spin
glass.  These results indicate that for the case in which the bonds
are chosen randomly to have values of $\pm J$ conformal
invariance\cite{AHHM06} and gauge invariance\cite{Tou77,HW05}
combine to create a set of remarkable properties.  In this work we
will present more such remarkable properties of the model, and find
a numerical result which seems to rule out some of these proposals.

The Hamiltonian of the Edwards-Anderson spin-glass model\cite{EA75}
for Ising spins is
\begin{equation}
  H = - \sum_{\langle ij \rangle} J_{ij} \sigma_{i} \sigma_{j}   \, ,
\end{equation}
where each spin $\sigma_{i}$ is a dynamical variable which has two
allowed states, $+1$ and $-1$.  The $\langle ij \rangle$ indicates a
sum over nearest neighbors on a simple square lattice of size $L
\times M$. We choose each bond $J_{ij}$ to be an independent
identically distributed quenched random variable, with the
probability distribution
\begin{equation}
  P ( J_{ij} ) = p \delta (J_{ij} + 1)~+~(1 - p) \delta (J_{ij} -
  1)   \, ,
\end{equation}
so that we actually set $J = 1$, as usual.  Thus $p$ is these
concentration of antiferromagnetic bonds, and $( 1 - p )$ is the
concentration of ferromagnetic bonds.  In this work we will study
the case $p = 0.5$, for which the average of $P ( J_{ij} )$ is zero.

In two dimensions (2D), the spin-glass phase is not stable at finite
temperature.  Because of this, it is necessary to treat cases with
continuous distributions of energies (CDE) and cases with quantized
distributions of energies (QDE) separately.\cite{BM86,AMMP03}

Amoruso, Hartmann, Hastings and Moore\cite{AHHM06} have proposed
that in 2D there is a relation
\begin{equation}
  d_S = 1 + {3 \over {4 ( 3 + \theta_E )}}   \, ,
\end{equation}
where $d_S$ is the fractal dimension of domain walls, and $\theta_E$
is the exponent which characterizes the scaling of the domain wall
energy, $E_{dw}$, with size.  For the CDE case, the existing
numerical estimates\cite{AHHM06,BLM06} of $d_S$ and $\theta_E$
satisfy Eqn.~(3).

For the QDE case, it is known that $\theta_E = 0$.\cite{AMMP03,HY01}
Using Eqn.~(3) then gives $d_S = 1.25$.  The derivation of Eqn.~(3)
assumes that the critical exponent $\eta$ for the scaling of
spin-glass correlations is equal to zero, however.  This appears to
fail in the QDE case.\cite{KL05,PB05,Har07}

As pointed out by Wang, Harrington and Preskill,\cite{WHP03} domain
walls of zero energy which cross the entire sample play a special
role when the boundary conditions are periodic and/or antiperiodic
in both directions and the energy is quantized. In the work
presented here, we will find that the properties of these $E_{dw} =
0$ domain walls are very special indeed.

\section{Numerical results for ground state entropy}

We will analyze data for the domain wall entropy, $S_{dw}$, for the
ground states (GS) of 2D Ising spin glasses obtained using a
slightly modified version of the computer program of Gallucio, Loebl
and Vondr\'{a}k,\cite{GLV00,GLV01} which is based on the Pfaffian
method.  The Pfaffians are calculated using a fast exact integer
arithmetic procedure, coded in C++.  Thus, there is no roundoff
error in the calculation until the double precision logarithm is
taken to obtain $S_{dw}$.  This extended precision is essential, in
order to obtain meaningful results for entropy differences for large
values of $L$ and $M$.  An earlier version of this
calculation\cite{Fis06b} was limited to $L \times L$ lattices.

We define the GS entropy to be the natural logarithm of the number
of ground states.  For each sample the GS energy and GS entropy were
calculated for the four combinations of periodic (P) and
antiperiodic (A) toroidal boundary conditions along each of the two
axes of the square lattice.  We will refer to these as PP, PA, AP
and AA.  The computer program treats the two lattice directions on
an equal footing.  Therefore we assume, without loss of generality,
that $L \le M$.

\begin{figure}
\includegraphics[width=3.4in]{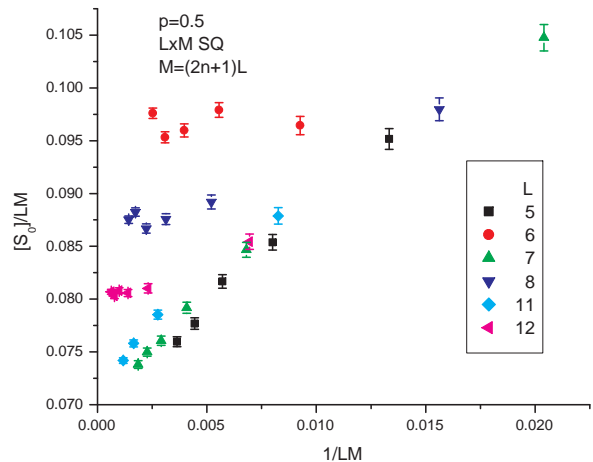}
\caption{\label{Fig.1}(color online) Average GS entropy per spin, $[
S_0 ] / L M$ vs. $1 / L M$, for $L \times M$ lattices. The error
bars indicate one standard deviation statistical errors.}
\end{figure}

For each lattice size $L \times M$ which was studied, 500 samples of
the random bonds were used to calculate statistical averages for
quantities of interest.  In Fig.~1 we show the average GS entropy $[
S_0 ( L , M ) ]$ per spin, where the brackets [~~] indicate an
average over random samples of the $J_{ij}$, for a large number of
sizes $L \times M$. The values of $M$ shown here are chosen to be $2
n + 1$, where $n$ is 0, 1, 2 ...~.  Thus for these lattices $L$ and
$M$ are either both odd or both even.  We see that the behaviors for
odd $L$ and even $L$ are distinct.  When the aspect ratio, $R = M /
L$ becomes large, $[ S_0 ( L , M ) ]$ for the lattices with odd $L$
approaches $S_0 ( \infty , \infty )$ from below, while for even $L$
it approaches this limit from above.  $R$ functions as a control
parameter which takes us from the $L \times L$ systems, for which $[
S_0 ]$ falls on a single curve for both odd $L$ and even
$L$,\cite{Fis06b} to the $L \times \infty$ systems, for which $[ S_0
]$ has distinct behaviors for odd $L$ and even $L$.

We define domain walls for the spin glass as it was done in the
seminal work of McMillan.\cite{McM84b}  We look at differences
between two samples with the same set of bonds, and the same
boundary conditions in one direction, but different boundary
conditions in the other direction.  Thus, for each set of bonds we
obtain domain wall data from the four pairs (PP,PA), (PP,AP),
(AA,PA) and (AA,AP).  For each size $L \times M$ we have 1000 data
points for the short (horizontal) direction, and another 1000 data
points for the long (vertical) direction.

The domain-wall renormalization group\cite{McM84a} is based on the
idea that we are studying an effective coupling constant which is
changing with $L$ and $M$.  For the CDE case\cite{CBM02} we can use
the domain wall energy, $E_{dw}$, which is defined to be the change
in the GS energy when the boundary condition is changed along one
direction from P to A (or vice versa), with the boundary condition
in the other direction remaining fixed, as the coupling constant.
For the QDE case, what we need to do is a slight generalization of
this idea. We should think of the coupling constant as the free
energy at some infinitesimal temperature.  When we do this, the
entropy contributes to the coupling constant.

The domain wall entropy, $S_{dw}$, is defined\cite{Fis06b} to be the
change in $S_0$ when the boundary condition is changed along one
direction from P to A (or vice versa), with the boundary condition
in the other direction remaining fixed.  As long as $E_{dw} > 0$,
the two boundary conditions which we are comparing are not on an
equal footing.  At a fixed aspect ratio, $[S_{dw}]$ is expected to
increase as a positive power of $L$ for any $E_{dw} > 0$. Therefore,
these coupling constants must eventually, at large enough $L$, be
controlled by $[S_{dw}]$ for any $T > 0$. Of course, the value of
$L$ which is needed for this to happen depends in $T$.

As Wang, Harrington and Preskill\cite{WHP03} express the situation,
an $E_{dw} > 0$ domain wall does not destroy the topological
long-range order.  However, in the $E_{dw} = 0$ case the two
boundary conditions are on an equal footing, and the topological
order is destroyed.  The probability distribution of $S_{dw}$ for
the cases where $E_{dw} = 0$ should be symmetric about 0, and our
statistics are consistent with this. Therefore the $E_{dw} = 0$
class of domain walls can be expected to behave in a special way.

It is important to realize that the meaning of a domain wall is very
different when $S_0$ is positive, as in the model we study here, as
compared to the typical case of a doubly degenerate ground state. In
the case of two-fold degeneracy one can identify a line of bonds
which forms a boundary between regions of spins belonging to the two
different ground states.  It is not possible, in general, to do that
when there are many ground states.  Despite this, we continue to use
the term ``domain wall".

\section{$S_{dw}$ for even $L$}

\begin{figure}
\includegraphics[width=3.4in]{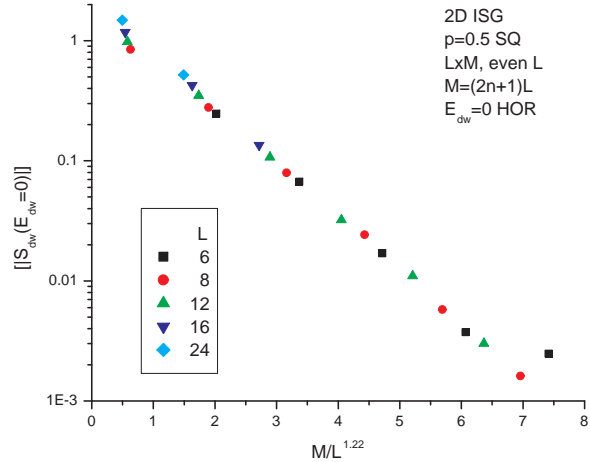}
\caption{\label{Fig.2}(color online) Finite-size scaling function
for $[ |S_{dw}( L, M )| ]$ vs. $M / L^{1.22}$ for $E_{dw} = 0$
domain walls which run in the $L$ direction. The $y$-axis is scaled
logarithmically.}
\end{figure}

When $L$ is even, the energy difference, $E_{dw}$, for any pair of
GS for which the boundary conditions are changed in the horizontal
direction, with the boundary conditions in the vertical direction
remaining fixed, must be a multiple of 4. When $L$ is odd and the
boundary conditions are changed in this way, $E_{dw}$ is $4 n + 2$,
where $n$ is an integer.\cite{BM86,Fis06b}  Equivalent statements
are true for odd and even $M$, with the roles of the horizontal and
vertical boundary conditions interchanged. The sign of $E_{dw}$ for
a McMillan pair is essentially arbitrary for $p = 1/2$. Thus we can,
without loss of generality, choose all of the domain-wall energies
to be non-negative.

The probability distribution for $E_{dw}$ is a strong function of
the aspect ratio of the lattice.  When $L$ is even the probability
that $E_{dw} \ne 0$ goes exponentially to 0, as a function of
$R$.\cite{HBCMY02}  Similarly when $L$ is odd we find that the
probability that $E_{dw} \ne 2$ goes exponentially to zero as $R$
increases.  However, the difference between even $L$ and odd $L$ for
large $R$ turns out to be profound.

For the $E_{dw} = 0$ case it is convenient to study $| S_{dw} |$,
since the distribution is symmetric about zero.  In Fig.~2 we
display a scaling function for the behavior of $[ | S_{dw} | ]$
versus $M / L^{d_S}$, for domain walls which run in the $(L)$
direction, with
\begin{equation}
  d_S = 1.22 \pm 0.01   \, ,
\end{equation}
where the error estimate is a one-standard-deviation statistical
error.  This estimate of $d_S$ is obtained from comparing the
behavior of the $L = 8$ data and the $L = 12$ data as a function of
$M$, and requiring that the slopes in Fig.~2 should be identical.
Thus we find that the true value of $d_S$ is probably less than
1.25, and the possibility that $d_S$ is the same for the QDE case as
the CDE case,\cite{JLMM06} for which $d_S$ is approximately
1.27,\cite{Fis07b} appears to be ruled out.

The naive choice of scaling variable, $R = M / L$, fails to produce
a satisfactory data collapse for data with differing values of $L$.
$L^{d_S}$ is the effective length of a domain wall for this model.
It is intuitively reasonable that the domain wall entropy should be
proportional to the length of the domain wall, rather than $L$.
Although such data are not displayed here, we have found that when
$M > L$ the results for odd $M$ with even $L$ also fall on the
scaling curve shown in Fig.~2.

Recently, Melchert and Hartmann\cite{MH07} have attempted to
calculate the scaling exponent for the length of domain walls for
this model directly.  They were unable to determine a precise value,
however, due to the high ground state degeneracy and the fact that
their algorithm does not select a ground state randomly.  Similar
results for hexagonal lattices have been given by Weigel and
Johnston.\cite{WJ07}

It is remarkable that $[ | S_{dw} | ]$ is falling exponentially as a
function of the variable $M / L^{d_S}$.  There is no $L$-dependent
scale factor for the $y$-axis. This is the {\it total} $| S_{dw} |$
for the entire $L \times M$ lattice. This behavior indicates that
the zero-energy domain walls which encircle the lattice in the short
direction must become strongly correlated as $R$ becomes large.
Because the Hamiltonian does not contain any explicit long-range
interactions, the mechanism by which this occurs is not trivial to
understand.

\begin{figure}
\includegraphics[width=3.4in]{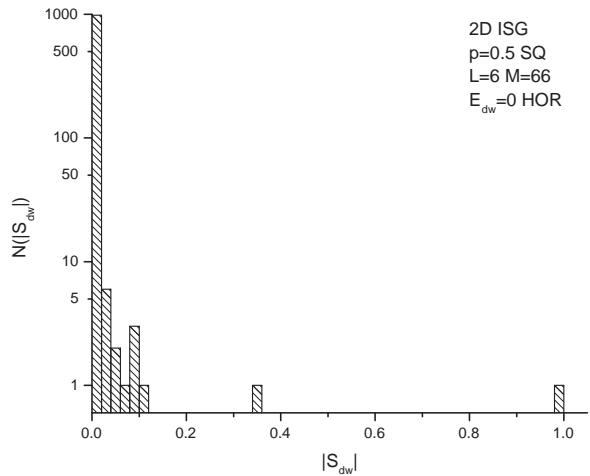}
\caption{\label{Fig.3} Histogram of the probability distribution of
$[|S_{dw}( 6, 66 )|]$ for $E_{dw} = 0$ domain walls which run in the
$L$ direction. The $y$-axis is scaled logarithmically.}
\end{figure}

\begin{figure}
\includegraphics[width=3.4in]{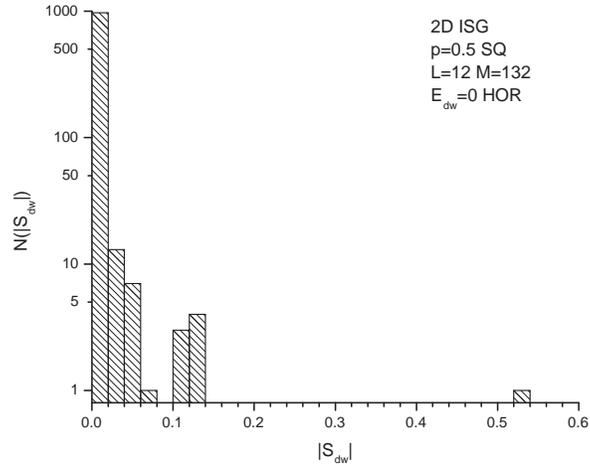}
\caption{\label{Fig.4} Histogram of the probability distribution of
$[|S_{dw}( 12, 132 )|]$ for $E_{dw} = 0$ domain walls which run in
the $L$ direction. The $y$-axis is scaled logarithmically.}
\end{figure}

As $M / L^{d_S}$ increases, the distribution of $[ | S_{dw} | ]$
becomes increasingly singular.  The last point shown in Fig.~2 is
above the trend because it is dominated by a single data point.  To
make this issue concrete, in Fig.~3 we show the histogram of the
probability distribution for this point, $| S_{dw}(6,66) |$. The
domination of the mean of this distribution by the point at the far
right of the histogram is obvious by inspection.  To demonstrate
that this is typical behavior, we show the corresponding histogram
for $| S_{dw}(12,132) |$, which is not off the trend line, in
Fig.~4.

As the reader may have already noticed, no statistical error bars
are given for the data points in Fig.~2.  In order to give a
meaningful estimate of such statistical errors for these probability
distributions, we would need to know what the analytical forms of
the probability distributions are.  We do not have this information.
The statistical error estimate for $d_S$ comes from fitting the
observed fluctuations of the data points from the trend line.  For
this we do not need to know the statistical errors of the individual
data points.

\begin{figure}
\includegraphics[width=3.4in]{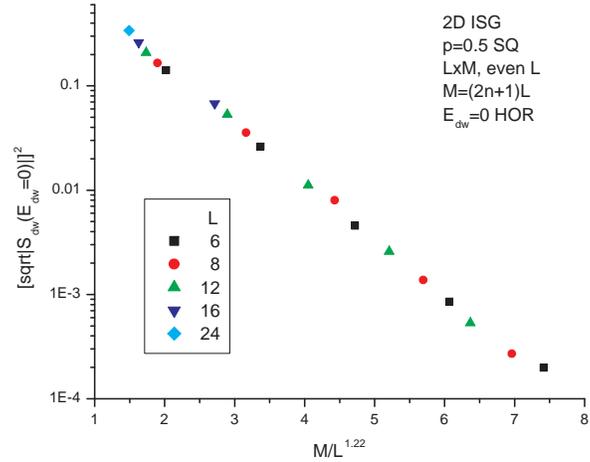}
\caption{\label{Fig.5}(color online) Finite-size scaling function
for $[\sqrt{ |S_{dw}( L, M )| }]^2$ vs. $M / L^{1.22}$ for $E_{dw} =
0$ domain walls which run in the $L$ direction. The $y$-axis is
scaled logarithmically.}
\end{figure}

Because the probability distributions of $| S_{dw} |$ become so
singular in the limit of large $R$, they appear to be
multifractal\cite{PV87}. One way of seeing this is to calculate the
fractional moments
\begin{equation}
  |S_{dw}(L,M)|_q = [|S_{dw}( L, M )|^{1/q}]^q   \, ,
\end{equation}
for $q$= 1, 2, 3 ... when $R > 1$ for the $E_{dw}=0$ domain walls
which run in the $L$ direction.  If $| S_{dw} |$ was controlled by a
single length scale, the slopes of the scaling functions for
$|S_{dw}(L,M)|_q$, analogous to the $q = 1$ case shown in Fig.~2,
would be identical (within statistical errors).  When one does this
calculation, however, one finds that the rate at which
$|S_{dw}(L,M)|_q$ decays exponentially to zero as $M / L^{d_S}$
increases is an increasing function of $q$.  The results for $q = 2$
are shown in Fig.~5.  It is clear that the data points are falling
faster as $M / L^{1.22}$ increases for $q = 2$ than they do for $q =
1$.  (Note that the scales on the axes of Fig.~5 are different from
those of Fig.~2.)  This trend continues for larger values of $q$.

The probability distributions for $| S_{dw} |$ of $E_{dw}=0$ domain
walls are thus inconsistent with any single-length scaling
hypothesis.  The author believes that this behavior arises from the
fact that the eigenvectors of the susceptibility matrix must be
Anderson-localized in 2D.\cite{And79}  Multifractal behavior is a
natural result of disorder-induced localization.\cite{PV87}

In an earlier work,\cite{Fis06b} the author calculated the
finite-size scaling behavior of $| S_{dw} |$ for $E_{dw}=0$ domain
walls on $L \times L$ square lattices. It was found there that the
exponent $\theta_S$ for the scaling of the width of the probability
distributions for $| S_{dw} |$ depends on $E_{dw}=0$.  For
$E_{dw}=0$ domain walls
\begin{equation}
  \theta_S ( E_{dw} = 0 ) = 0.500 \pm 0.020   \, .
\end{equation}
According to the droplet model, one expects $d_S = 2
\theta_S$.\cite{FH88}  The value of $d_S$ which we have found here
is clearly inconsistent with this relation for $E_{dw} = 0$. The
reason for this is the special symmetry of the $E_{dw} = 0$ domain
walls, as discussed in the earlier work.\cite{Fis06b}

\section{$S_{dw}$ for odd $L$}

For odd $L$ with the boundary conditions we are using, $E_{dw}$
cannot be zero for domain walls which run in the $L$ direction.
When $E_{dw}$ is not zero, the relative signs of $E_{dw}$ and
$S_{dw}$ are not arbitrary. Having chosen $E_{dw}$ to be positive,
we then find that $S_{dw}$ is also usually positive.\cite{Fis06b} In
Fig.~6 we show the finite-size scaling of $[ S_{dw}( L, M ) ]$ for
the $E_{dw} = 2$ domain walls which run in the $L$ direction. In
order to make the data for different values of $L$ fall on a common
curve, it is necessary to include a $y$-axis scale factor of $L^{-1
/ 2}$. The scaling variable again appears to be $M / L^{d_S}$, with
the same value of $d_S$ as before, although our uncertainty in the
value of $d_S$ is larger for odd $L$.  The behavior seems to be a
power-law increase, with an slope close to $0.30$. The contrast of
this behavior with the exponential decrease which we found for the
$E_{dw} = 0$ walls could hardly be greater.

\begin{figure}
\includegraphics[width=3.4in]{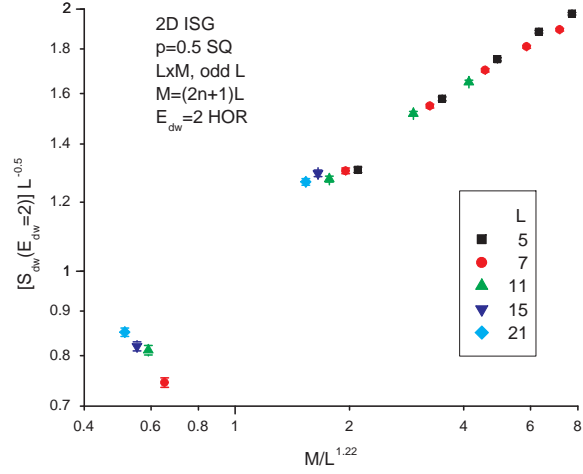}
\caption{\label{Fig.6}(color online) Finite-size scaling function
for $[ S_{dw}( L, M ) ] L^{-0.5}$ vs. $M / L^{1.22}$ for $E_{dw} =
2$ domain walls which run in the $L$ direction, log-log plot.}
\end{figure}

\begin{figure}
\includegraphics[width=3.4in]{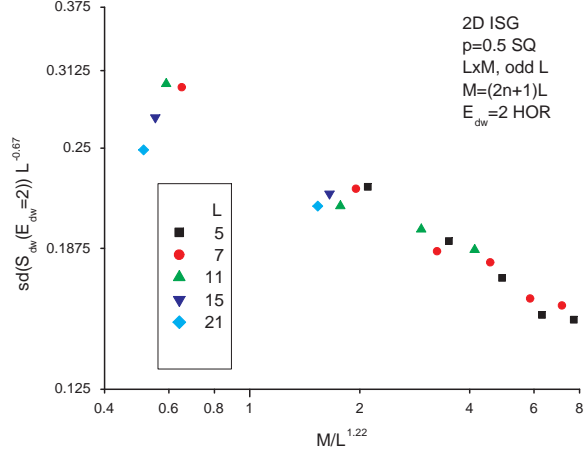}
\caption{\label{Fig.7}(color online) Finite-size scaling function
for the standard deviation, $sd( S_{dw}( L, M ) ) L^{-0.67}$ vs. $M
/ L^{1.22}$ for $E_{dw} = 2$ domain walls which run in the $L$
direction, log-log plot.}
\end{figure}

In Fig.~7 we show the finite-size scaling behavior for the widths of
the same $E_{dw} = 2$ $S_{dw}( L, M )$ distributions, as
parameterized by their standard deviations.  The $y$-axis scale
factor for data collapse appears to be $L^{-2 / 3}$ in this case.
The behavior again appears to be a power-law in $M / L^{d_S}$.  The
slope in this case is approximately $-0.30$.  Thus the widths of
these $S_{dw}( L, M )$ distributions should eventually become small
compared to their average values as $R$ increases.

For domain walls that run in the long, $M$ direction, the
probability that $E_{dw} = 0$ goes to zero as $R \to
\infty$.\cite{FH06}  We do not expect any anomalous behavior for the
``long" domain walls.  All of the results found here are consistent
with the conclusion\cite{Fis06b} that in this model there seem to be
two distinct classes of domain walls, the $E_{dw} = 0$ domain walls
and the $E_{dw} > 0$ domain walls.

The alert reader has noticed that the sizes of the lattices for
which we have data when $L$ is odd are more limited than in the even
$L$ case.  This is because the Vondr\'{a}k code runs about a factor
of four faster if both $L$ and $M$ are even, due to symmetry
properties of the Pfaffians.

\section{Summary}

We have studied the statistics of domain walls for ground states of
the 2D Ising spin glass with $+1$ and $-1$ bonds for $L \times M$
square lattices with $p$ = 0.5, where $p$ is the fraction of
negative bonds, using periodic and/or antiperiodic boundary
conditions, for both even and odd $L$ and $M$, where $L \le M$.  The
probability distributions of domain wall entropy, $S_{dw} (L,M)$,
are found to depend strongly on $E_{dw}$, and therefore on whether
$L$ is odd or even.  Finite-size scaling forms are found which are
functions of the variable $M / L^{d_S}$, where $d_S = 1.22 \pm
0.01$. When the aspect ratio becomes large, the distribution of
$S_{dw}$ for zero-energy domain walls which encircle the lattice in
the short direction becomes multifractal.

\begin{acknowledgments}
The author thanks J. Vondr\'{a}k for providing a copy of his
computer code, and for help in learning how to use it.  He is
grateful to A. K. Hartmann, D. A. Huse and M. A. Moore for helpful
discussions, and to the Physics Department of Princeton University
for providing use of the computers on which the data were obtained.

\end{acknowledgments}


\begin{thebibliography}{49}
\bibitem{JLMM06}
T. J\"{o}rg, J. Lukic, E. Marinari and O. C. Martin, Phys. Rev.
Lett. {\bf 96}, 237205 (2006).
\bibitem{AHHM06}
C. Amoruso, A. K. Hartmann, M. B. Hastings and M. A. Moore, Phys.
Rev. Lett. {\bf 97}, 267202 (2006).
\bibitem{Fis06b}
R. Fisch, J. Stat. Phys. {\bf 125}, 793 (2006).
\bibitem{Fis07}
R. Fisch, J. Stat. Phys. {\bf 128}, 1113 (2007).
\bibitem{Tou77}
G. Toulouse, Commun. Phys. {\bf 2}, 115 (1997).
\bibitem{HW05}
M. B. Hastings and X.-G. Wen, Phys. Rev. B {\bf 72}, 045141 (2005).
\bibitem{EA75}
S. F. Edwards and P. W. Anderson, J. Phys. F {\bf 5}, 965 (1975).
\bibitem{BM86}
A. J. Bray and M. A. Moore, in {\it Heidelberg Colloquium on Glassy
Dynamics}, J. L. van Hemmen and I. Morgenstern, ed., (Springer,
Berlin, 1986), pp. 121-153.
\bibitem{AMMP03}
C. Amoruso, E. Marinari, O. C. Martin and A. Pagnani, Phys. Rev.
Lett. {\bf 91}, 087201 (2003).
\bibitem{BLM06}
D. Bernard, P. Le Doussal and A. A. Middleton, Phys. Rev. B {\bf
76}, 020403(R) (2007).
\bibitem{HY01}
A. K. Hartmann and A. P. Young, Phys. Rev. B {\bf 64}, 180404(R)
(2001).
\bibitem{KL05}
H. G. Katzgraber and L. W. Lee, Phys. Rev. B {\bf 71}, 134404
(2005).
\bibitem{PB05}
J Poulter and J. A. Blackman, Phys. Rev. B {\bf 72}, 104422 (2005).
\bibitem{Har07}
A. K. Hartmann, arXiv:0704.2748.
\bibitem{WHP03}
C. Wang, J. Harrington and J. Preskill, Ann. Phys. (N.Y.) {\bf 303},
31 (2003).
\bibitem{GLV00}
A. Galluccio, M. Loebl and J. Vondr\'{a}k, Phys. Rev. Lett. {\bf
84}, 5924 (2000).
\bibitem{GLV01}
A. Galluccio, M. Loebl and J. Vondr\'{a}k, Math. Program., Ser. A
{\bf 90}, 273 (2001).
\bibitem{McM84b}
W. L. McMillan, Phys. Rev. B {\bf 29}, 4026 (1984).
\bibitem{McM84a}
W. L. McMillan, J. Phys. C {\bf 17}, 3179 (1984).
\bibitem{CBM02}
A. C. Carter, A. J. Bray, and M. A. Moore, Phys. Rev. Lett. {\bf
88}, 077201 (2002).
\bibitem{HBCMY02}
A. K. Hartmann, A. J. Bray, A. C. Carter, M. A. Moore and A. P.
Young, Phys. Rev. B {\bf 66}, 224401 (2002).
\bibitem{Fis07b}
R. Fisch, arXiv:0705.0046.
\bibitem{MH07}
O. Melchert and A. K. Hartmann, arXiv:0704.2004.
\bibitem{WJ07}
M. Weigel and D. Johnston, Phys. Rev. B {\bf 76}, 054408 (2007).
\bibitem{PV87}
G. Paladin and A. Vulpiani, Phys. Rev. B {\bf 35}, 2015 (1987).
\bibitem{And79}
P. W. Anderson, in {\it Ill-Condensed Matter}, R. Balian, R. Maynard
and G. Toulouse, ed., (North-Holland, Amsterdam, 1979), pp. 214-261.
\bibitem{FH88}
D. S. Fisher and D. A. Huse, Phys. Rev. B {\bf 38}, 386 (1988).
\bibitem{FH06}
R. Fisch and A. K. Hartmann, Phys. Rev. B {\bf 75}, 174415 (2007).


\end{thebibliography}


\end{document}